\documentstyle[floats,prd,aps,twocolumn]{revtex}
\draft 

\def\be{\begin{equation}}
\def\ee{\end{equation}}
\def\bea{\begin{eqnarray}}
\def\eea{\end{eqnarray}}
\def\l{\label}
\def\th{\theta}
\def\la{\lambda}
\def\al{\alpha}

\begin{document}

\title{ Cosmological model in 5D, stationarity, yes or no }

\author{ W. B. Belayev\thanks{Electronic address: vladter@ctinet.ru}}

\address{ Center for Relativity and Astrophysics, box 137, 194355, Sanct-Petersburg, 
Russia }

\maketitle

\begin{abstract}
We consider cosmological model in 4+1 dimensions with variable scale factor in extra 
dimension and static external space. The time scale factor is changing. Variations of light 
velocity, gravity constant, mass and pressure are determined with four-dimensional 
projection of this space-time. Data obtained by space probes Pioneer 10/11 and Ulysses are 
analyzed within the framework of this model. 
\end{abstract}


\section{INTRODUCTION}

Theory, explaining the redshift of the spectra of distant galaxies with variation of time scale 
factor was put forth by Milne \cite{milne1}. He also thought the dependence of gravity 
constant on time $t$ to be possible \cite{milne2}. Dirac \cite{dirac} considered gravity 
constant to be inversely proportional to $t$, and time dependence of other fundamental 
constants. However, with the constant length scale factor these theories can not explain the 
proximity of the density of matter in the Universe to critical value, which follows from 
Fridman-Robertson-Walker model. At the same time, according to different 
estimates\cite{bahc} the cosmological density does not exceed 1/3 of this value. 

In this connection Kaluza-Klein theory is of interest. A review of articles about this theory is 
included in \cite{overd}. Internal space is considered to be forming an extra dimension in 
five-dimensional space-time \cite{overd,wesson}. Models with extra dimensions and 
possibilities of variation of existing constants are analyzed in \cite{kolb,barrow}. Variation 
of bare "constants" of nature might be caused by variation of scale factor of internal spaces 
$R$ . Though presented theories give estimate of rate of relative changing of $R$ orders of 
magnitude smaller than Hubble constant $H$, as pointed out  Barrow \cite{barrow}, they 
base on variation of only several constants, leaving other unchanged. Operating on the 
principle of similarity of processes, having a theory that variation of constants is determined 
by metric properties of space, one may have another result. However, this assumption needs 
an additional reasoning.

Dependencies of light velocity $c$, Planck constant, energy of a particle, its mass, 
magnitude of force in various coordinate systems in space-time of Minkowsky with cosmic 
time on scale factor of time $N$  are analyzed in Sec. \ref{sec21}. Redshift dependence of 
$N$ and correlation between variation of $c$ and $H$ of given space-time are determined 
in Sec. \ref{sec3}. Sec. \ref{sec4} contains analysis of Schwarzschild's metric with cosmic 
time, presenting dependence of gravitational constant and body motion in gravitational field 
on $N$. In Sec. \ref{sec5} energy processes caused by the change of time scale factor are 
investigated. Radiometric data from Pioneer 10/11 and Ulysses spacecraft are analyzed in 
Sec. \ref{sec6}. Sec. \ref{sec7} contains solution of Einstein equations for five dimensions 
$( 4+1)$, giving possibility to create a cosmological model with length scale factor which is 
constant in three-dimensional space and variable in additional space with cosmic time. In 
Sec. \ref{sec8} the magnitude of critical density of matter in the Universe corresponding to 
this model is determined and dependence of redshift on distance is considered. 

\section{MINKOWSKY'S SPACE-TIME WITH COSMIC TIME}
\l{sec2}

\subsection{Mechanics} 
\l{sec21}

Let's consider a metric with the line element in four-dimensional space 
\be\l{f1}
ds^{2}=c_{0}^2N^2(t)dt^2-dr^2+r^2d\th^2+r^2sin^2\th d\phi^2,
\ee
where  $r$, $\th$, $\phi$ are the spherical coordinates, $c_{0}$ is the light velocity in a 
given moment of time $t_{0}$, $N$ is the coefficient dependent of $t$. Assuming $N(t)$ to 
be slow changing and making the substitution 
\be\l{f2}
c=c_{0}N(t)
\ee
we get the standard Minkowsky metric. Therefore, light velocity in four-dimensional 
coordinate system with its center in $(t,O)$, where $O$ is an arbitrary point moving without 
action of any forces  in three-dimensional space is represented by (\ref{f2}). Let us denote 
\be\l{f3}
\tau=\int_{0}^{t}N(t)dt
\ee
as a time in $(t_{0},O)$ coordinate system. The expression tied a distance intervals 
$dr_{0}$  in $(t_{0},O)$ system and $dr$ in $(t,O)$ system follows from metrics 
(\ref{f1}):
\be\l{f4}
dl_{0}=dl.
\ee
Then, a velocity $v_{0}$ in $(t_{0},O)$ coordinate system could be represented as 
\be\l{f5}
v_{0}=\frac{dl_0}{d\tau}.
\ee

The energy of a particle for the metric (\ref{f1}) is \cite{land}:
\be
E=\frac{m_{0}c_{0}^{2}N(t)}{\left(1-v_{0}^2/c_{0}^2\right)^{\frac{1}{2}}},
\ee
where $m_{0}$ is the rest mass of the particle in $(t_{0},O)$ coordinate system. Assuming 
$v_{0}=0$, we get expression for rest energy of the particle 
\be\l{f6}
E=m_{0}c_{0}^{2}N(t).
\ee
Hence
\be\l{f7}
E=E_{0}N(t),
\ee
where $E_{0}=m_{0}c_{0}^2$ is the rest energy of the particle in coordinate system with 
its center in $(t_{0},O)$. Detailed discussion of the result (\ref{f7}) will be given in Sec. 
\ref{sec5}. In view of (\ref{f2}) we obtain from (\ref{f6}) $E=m_{0}c^2/N(t)$. It follows 
herefrom that the rest mass is changing with time 
\be\l{f8}
m(t)=\frac{m_0}{N(t)}.
\ee
This variation of mass is relative, i.e, does not occur as variation of mass as number of 
nucleons. 

Let us determine the magnitude of force acting upon the particle. Expression for the vector of 
the force  at small velocities in relation to coordinate system $(t,O)$ is represented by 
$F=mdV/dt,$ where $V$ is the velocity vector. Since (\ref{f5}) the velocity vector in 
coordinate system $(t_{0},O)$ is $V_{0}=V/N(t),$ because of (\ref{f8}), (\ref{f3}) the 
vector of the force is written as
\be
F=\frac{m_0}{N(t)}\frac{d\left(V_{0}N(t)\right)}{dt}=\frac{m_0}{N(t)}\left[\dot{N}(t)V
_{0}+ \frac{dV_0}{d\tau}N^{2}(t)\right],
\ee
where and below overdot denote the derivative with respect to $t.$ Regarding $\dot{N}$ to 
be small, we get the magnitude of force, acting upon the particle in system $(t,O):$
\be\l{f9}
f=m_{0}\frac{dv_0}{d\tau}N(t)=f_{0}N(t),
\ee
where $f_0$ is its value in system $(t_{0},O)$.

Let us consider variation of Planck constant $h$ with time. The quantum energy in system 
$(t,O)$ is $E=h/P,$ where $P$ is the radiation period. Considering the energy of quantum 
changing in accordance with (\ref{f7}) we will obtain, taking into account (\ref{f3}):
\be
E_{0}N(t)=\frac{hN(t)}{P_{0}(\tau)},
\ee
where $P_{0}(\tau)$ is the radiation period, which is correspondent to energy of quantum, 
emitted in the moment $\tau,$ within coordinate system $(t_{0},O).$ As long as 
$E_{0}=h_{0}/P_{0},$ where $h_0$ is the Planck constant, and $P_0$ is the emission 
period in system $(t_{0},O),$ we may assume $h$ to be constant
\be
h=h_{0}.
\ee

\subsection{Cosmological parameters}
\l{sec3}

We determine the redshift magnitude for the cosmological model with the metric (\ref{f1}):   
\be
z=\frac{\left(\la_{0}(\tau)-\la_{0}(\tau_{0})\right)}{\la_{0}(\tau_{0})},
\ee
where $\la_{0}(\tau)$ is the wavelength of emission, radiated in moment $\tau$ in system 
$(t_{0},O),$ $\tau_0= \tau(t_0).$  This formula gives 
\be 
z=\frac{\left(P_{0}(\tau)-P_{0}(\tau_{0})\right)}{P_{0}(\tau_{0})}.
\ee
Since $P_{0}(\tau)=PN(t)$ and the period of emission $P$ is constant at any system, bound 
with the emission time: $P=P_{0}(\tau_{0})$, then in view of $N(t_0)=1$ we have
\be\l{f10}
z=N(t)+1.
\ee
Assuming the rate of change $N(t)$ to be constant in time interval $\Delta t=t_{0}-t$, we 
write $z=-\dot{N}(t_0)\Delta t_{0}.$ Denoting
\be\l{f11}
H=-\frac{\dot{N}(t_0)}{N(t_0)},
\ee
we obtain Hubble Law $zc_{0}=H\Delta r_0,$ where $\Delta r_{0}=\Delta t_{0}c_0.$

Let us determine the variation of light velocity per time unit from (\ref{f2}):
\be\l{f12}
\dot{c}(t_0)=c_{0}\dot{N}(t_0)=-c_{0}H
\ee
Assuming the Hubble constant $H=50$ $\rm{km\;s^{-1}\;Mpc^{-1}}$  \cite{good,sand},  
we will get $\dot{c}(t_0)=1.5$ $\rm{cm\;s^{-1}\;yr^{-1}}.$

Now we consider the temperature change of microwave cosmic background. Since the 
energy of a quantum of light changes in accordance with (\ref{f7}), the temperature is
\be
T=T_{0}N(t),
\ee
where $T_0$ is the temperature of microwave cosmic background at present.

\section{SCHWARZCHILD'S METRIC WITH COSMIC TIME}
\l{sec4}

Let us consider the metric
\bea
ds^2=\left(1-\frac{\al}{r}\right)N^2(t)dt^2 \nonumber \\
-\frac{1}{c_{0}^2}\left[\left(1-\frac{\al}{r}\right)^{-1}dr^2+ r^2d\th^2+r^2sin^2\th 
d\phi^2\right],
\eea
where $\al$ is the constant. Assuming $N(t)$ slowly changing with $t$ and the movement to 
be radial, we solve the equations of the geodesic line \cite{vittie}:
\bea\l{f13}
\frac{d}{ds}\left[\left(1-\frac{\al}{r}\right)N^2\frac{dt}{ds}\right]=0 \\
\l{f14}
\left(1-\frac{\al}{r}\right)N^2\left(\frac{dt}{ds}\right)^2-\frac{1}{c_{0}^2}\left(1-
\frac{\al}{r}\right)^{-1}\left(\frac{dr}{ds}\right)^2 =1.
\eea
Integrating (\ref{f13}) we obtain 
\be
\frac{dt}{ds}=\beta \left(1-\frac{\al}{r}\right)^{-1}N^{-2},
\ee
where $\beta$ is the integration constant. We substitute this equation into (\ref{f14}): 
\be\l{f15}
\left(1-\frac{\al}{r}\right)^{-1}N^{-2}-\frac{1}{c_{0}^2}\left(1-\frac{\al}{r}\right)^{-
3}N^{-4}\left(\frac{dr}{dt}\right)^2 =\beta^{-2}.
\ee
If movement is such that particle can reach infinite values, then 
\be
\beta^{-2}=N^{-2}-\frac{1}{c_{0}^2}N^{-4}v_{i}^2,
\ee
where $v_{i}^2$ is the velocity of particle at infinity. Assuming $v_{i}^2N^{-2}\ll 
c_{0}^2,$ we  get $\beta^2=N^2.$ Then, equation (\ref{f15}) is written as
\be
\left(\frac{dr}{dt}\right)^2=c_{0}^{2}\left(1-\frac{\al}{r}\right)^{2}\frac{\al}{r}N^2.
\ee
Differentiating this expression by $t$ and taking into account small magnitude of $\al/r$ and 
slow variation of $N(t)$, we obtain 
\be\l{f16}
\frac{d^2r}{dt^2}=-\frac{\al c_{0}^{2}N^2}{2r^2}.
\ee
At $t=t_0$ this equation express Newton's Law for radial movement of particle in external 
gravitational field of a spherical mass, if $\al=2G_{0}M_{0}/c_{0}^2,$ where $G_0$ is 
the gravity constant in system $(t_0,O),$ $M_0$ is spherical mass in this coordinate system. 
Thus, (\ref{f16}) is presented as 
\be\l{f17}
\frac{d^2r}{dt^2}=-\frac{G_{0}M_{0}N^2}{2r^2}.
\ee
Because of (\ref{f8}) $M=M_{0}/N$ the expression, determining variation of gravity 
constant $G$ with time in formula for Newton's Law $d^2r/dt^2=-GM/r^2,$ is given by
\be\l{f18}
G=G_{0}N^{3}.
\ee
Now we change to $(t_{0},O)$ coordinate system in expression (\ref{f17}). Since values of 
distance and time intervals in this system are 
$dr_0=dr$ and $d\tau=N(t)dt,$ we have
\be
\frac{d^2r}{dt^2}=-
\frac{d}{dt}\left(\frac{dr_0}{d\tau}N(t)\right)=N^2(t)\frac{d^{2}r_{0}} 
{d\tau^2}+\dot{N}(t)\frac{dr_0}{d\tau}.
\ee
Assuming second term in the right side of equation to be small, we get
\be\l{f19}
\frac{d^2r_0}{d\tau^2}=-\frac{G_{0}M_{0}}{2r_{0}^2}.
\ee
Since right side of this formula determines force acting upon a body with unity mass at the 
distance $r_0$, orbit of the body, moving in external gravity field does not change within the 
$(t_0,O)$ coordinate system. Absence of dependence of a length element on choice of time 
zero means constancy of shape and size of the orbit. 

\section{ENERGY PROCESSES AND GENERATION OF MATTER}
\l{sec5}

We define the energy change (\ref{f6}) with time
\be\l{f20}
W=\frac{dE}{dt}=m_{0}c_{0}^{2}\dot{N}(t).
\ee
In view of (\ref{f2}) and (\ref{f8}) we can write $W=\frac{\dot{N}}{N}mc^{2},$
and, as follows from (\ref{f11}), at present time
\be\l{f21}
W=-Hm_{0}c_{0}^2.
\ee
Thus, energy of a body with mass $1$ $\rm{kg}$ decreases by $0.14$ $\rm J$ every 
second. Since, as follows from energy conservation law, it does not disappear, it means that 
the energy is liberated in some way. This conclusion, so it should seem, rules out the theory 
considered. In truth, according to (\ref{f21}) the Sun should liberate $2.8\times10^{29}$ 
$\rm{J/sec}$, while it actually does $3.8\times10^{26}$ $\rm{J/sec}$.

However, Bondi and Gold \cite{bondi}, Hoyle \cite{hoyle}, Jordan \cite{jordan} have put 
forth a hypothesis of generation of matter within the framework of expanding Universe 
model. Analysis of currents of helium, coming out of depths of the Earth, testifies about its 
radiogenic origin \cite{geoph}. On the basis of these data, the hypothesis of generation of 
matter within the Earth gains further development \cite {socol}. However, to prove this 
theory, extra arguments are  necessary.  

Let's determine, what quantity of matter could emit in the form of additional nucleons in a 
unit of time, if all energy (\ref{f20}) is used for its generation. We consider the change of 
mass $m_{0}(t)$ resulting from generation of matter in moment of time $\tau(t)$ in 
coordinate system $(t_0,O).$ This mass emits per one time unit (\ref{f20}):
\be
U=-W=-\dot{N}(t)m_{0}(t)c_{0}^2.
\ee
On the other hand, formula (\ref{f6}) yields 
\be
U=\dot{m}_{0}(t)N(t)c_{0}^2.
\ee
where $\dot{m}_{0}$ is the mass, generated per time unit in system $(t_0,O).$ Solving 
equation $\dot{m}_{0}(t)/m_{0}(t)=-\dot{N}(t)/N(t)$ with initial conditions 
$m_{0}(t_0)=1,$ $N(t_0)=1,$ we obtain
\be\l{f22}
m_{0}(t)=\frac{m_0}{N(t)}
\ee
and, at present time, 
\be\l{f23}
\dot{m}_{0}(t_0)=Hm_0.
\ee
It should be pointed out, however, it is not essential, that all emitted energy is used for 
generation of matter, but in several cases, probably, is released in some other forms, for 
instance, as heat. 

\section{ANALYSIS OF EXPERIMENTAL DATA}
\l{sec6}

During radiometric analysis of Pioneer 10/11 spacecraft's data \cite{anders}, additional 
acceleration  $a_r\approx 8.5\times 10^{-8}$ $\rm{cm/s^2},$ directed towards the Sun was 
detected. This value is on the verge of limits, determined by error estimate of observed 
acceleration of Ulysses \cite{anders}. According to result, obtained in Sec. \ref{sec4} 
(\ref{f19}), trajectory of a body does not change with time. We consider possible 
explanation of presence of quasi acceleration. Detecting change of the wavelength of a 
received wave $\la$ could be a resultant of two factors: decrease of the light velocity and 
variation of the cycle of radiation of the spacecraft generator $P_g$ in the scale of receiving 
equipment: 
\be
\dot{\la}=\dot{c}P_g+c\dot{P}_{g} \ .
\ee
Variation of the cycle of radiation could be related to dependence of processes, which 
determine work of generator and receiving equipment, on time scale factor. This problem 
needs further consideration. If we make assumption $\dot{P}_{g}(t_0)=-P_{g}(t_0)H,$ 
then in view of (\ref{f12}) we can write $\dot{\la}(t_0)=-2P_{g}(t_0)H.$ The rate of the 
relative change of wavelength is
\be\l{f25}
\frac{\dot{\la}(t_0)}{\la}=-2H.
\ee
At the same time  it is presented as $\dot{\la}/\la=a_{r}/c,$ in assumption  cycle of radiation 
and light velocity are constant, and variation of wavelength is caused by additional  
acceleration. Hence (\ref{f25}) one can get a value of $H=43.7$ $\rm{km\;s^{-1}\;Mpc^{-
1}},$  which is in agreement with its estimate \cite{reph,sand}.

 If  $P_g$ does not change, then the rate of the relative change of wavelength is
\be
\frac{\dot{\la}(t_0)}{\la}=-H,
\ee
This one gives a value of $H=87.4$ $\rm{km\;s^{-1}\;Mpc^{-1}}.$

\section{COSMOLOGICAL MODEL IN FIVE DIMENSIONS}

\subsection{Partial solution of Einstein's equations in five dimensions}
\l{sec7}

We consider the solution of the Einstein's equations in five dimensions \cite{overd}:  
\be\l{f26}
\hat R^{\al\beta}-\frac{1}{2}g^{\al\beta}\hat R=8\pi GT^{\al\beta},
\ee
where $T^{\al\beta}$ is the matter energy momentum tensor, $\hat R^{\al\beta}$ is the 
Ricci tensor, $\hat R$ is the scalar curvature of space, $g^{\al\beta}$ is the metric tensor 
components. We  assume that parameters such as gravity constant, mass, light velocity and 
pressure are slowly changing with time  in accordance with results, obtained in Sec. 
\ref{sec2}, \ref{sec4}. We look for solution in the form 
\be\l{f27}
ds^2=N^2(t,\psi)dt^2-
\frac{1}{c_{0}^{2}}\left[dx^{12}+dx^{22}+dx^{32}+R^{2}(t,\psi)d\psi^2\right],
\ee
where $x^1$, $x^2$, $x^3$ is the space coordinates of the four-dimensional  space-time, 
$\psi$ is the coordinate of extra dimension, created by compactified internal space, 
$N(t,\psi)$, $R(t,\psi)$ are the arbitrary functions $t$ and $\psi.$ This metrics reduces to 
(\ref{f1}) metrics at $\psi=const.$ Let's denote derivatives with respect to $\psi$ by $(*)$ 
and $x^{0}\equiv t$, $x^{4}\equiv\psi.$  So the nonzero components of a tensor $\hat 
R^{\al\beta}$ are 
\be
\begin{array}{cl}
\hat R^{00}=-\frac{\ddot{R}}{RN^4}+c_{0}^{2}\frac{N^{**}}{R^{2}N^{3}}+ 
\frac{\dot{N}\dot{R}}{RN^{5}}-c_{0}^2\frac{N^*R^*}{R^{3}N^3}, \\ \\
\hat R^{44}=- 
c_{0}^{4}\frac{N^{**}}{NR^4}+c_{0}^{2}\frac{\ddot{R}}{N^{2}R^{3}}+ 
c_{0}^{4}\frac{R^*N^*}{NR^{5}}-c_{0}^2\frac{\dot{R}\dot{N}}{N^{3}R^3}.
\end{array}
\ee
Value of the scalar curvature of the space turns out to be
\be
\hat R=g_{00}\hat R^{00}+g_{44}\hat R^{44}=-
2\frac{\ddot{R}}{RN^2}+2c_{0}^{2}\frac{N^{**}}{NR^{2}}.
\ee
The nonvanishing equations (\ref{f26}) are
\be\l{f28}
\begin{array}{cl}
\frac{\dot{N}\dot{R}}{RN^{5}}-c_{0}^2\frac{N^*R^*}{R^{3}N^3}=8\pi GT^{00}\\ \\
c_{0}^{4}\frac{R^*N^*}{NR^{5}}-c_{0}^2\frac{\dot{R}\dot{N}}{N^{3}R^3}=8\pi 
GT^{44}\\ \\
c_{0}^{2}\frac{\ddot{R}}{RN^2}-c_{0}^{4}\frac{N^{**}}{NR^{2}}=8\pi GT^{ii}, 
\text{ }i=1,2,3.
\end{array}
\ee
Hereinafter we will assume $N$ and $R$ independent from $\psi$:
\be\l{f31}
N^*=0,\text{ }R^*=0.
\ee

We consider a five-dimensional energy tensor with nonzero diagonal matrix elements  look 
like
\be
T_{i}^{j}:=
\left(\begin{array}{cl}
   \rho \ \ \ \ \ . \ \ . \ \ . \ \ .   \\ \\
   . \ \ \frac{p_{ext}}{c^2} \ \ . \ \ . \ \ .   \\ \\
   . \ \ . \ \ \frac{p_{ext}}{c^2} \ \ . \ \ .   \\ \\
   . \ \ . \ \ . \ \ \frac{p_{ext}}{c^2} \ \ .   \\ \\
   . \ \ . \ \ . \ \ . \ \ \frac{p_{int}}{c^2}   \\ 
\end{array}
\right) \ ,
\ee
where $\rho$ is the density of matter in the Universe, $p_{ext}$ is the isotropic pressure in 
external three-dimensional space, $p_{int}$ is the pressure in additional dimension. We 
write the energy tensor \cite{vittie,land}:
\be
T^{ij}=\left(\rho+\frac{p^{ij}}{c^2}\right)u^{i}u^{j}-g^{ij}\frac{p^{ij}}{c^2},
\ee
where $p^{ij}=0, \text{ }i\neq j;$ $p^{ii}=p_{ext},\text{ }i=0...3;$ $p^{44}=p_{int},$ 
and $u^{i}=dx^i/ds$ are the components of velocity vector of the fluid elements. Assuming 
\cite{vittie,einst}  $u^i=0,\text{ }i=1...4,$ from (\ref{f27}) we obtain  
\be
u^0=N^{-1}.
\ee
We present nonzero components of the energy tensor
\be\l{f32}
\begin{array}{cl}
T^{00}=\rho N^{-2} \\ \\
T^{ii}=c_{0}^{2}c^{-2}p_{ext}, \text{ }i=1,2,3 \\ \\
T^{44}=c_{0}^{2}c^{-2}p_{int}R^{-2}.
\end{array}
\ee

Now we change over to coordinate system $(t_0,O)$. Since an element of length does not 
change at that, in view of (\ref{f8}), $\rho=\rho_{0}(t)N^{-1},$ where $\rho_{0}(t)$ is the 
density of matter in $(t_0,O)$ system in time $\tau(t)$. A density $\rho_{0}(t)$ can 
decrease as a result of  transformation of  matter into electromagnetic radiation. When 
assuming a possibility of  generation of matter (\ref{f22}), its accretion is larger then its 
decrease in the result of  emission of energy, and therefore at $t\le t_0:$ 
\be
\rho_{0}N^{-1}(t)\le \rho_{0}(t)\le\rho_{0},
\ee
where $\rho_{0}\equiv\rho_{0}(t_0).$ From (\ref{f9}) follows that $p=p_{0}N(t),$ where 
$p$ and $p_0$ are the pressures in $(t,O)$ and $(t_0,O)$ coordinate systems accordingly. 
Thus, taking into account formula for the light velocity (\ref{f2}), the expressions 
(\ref{f32}) can be written as 
\be
\begin{array}{cl}
T^{00}=\rho N^{-3} \\ \\ 
T^{ii}=p_{0ext}N^{-1}, \text{ }i=1,2,3 \\ \\
T^{44}=p_{0int}N^{-1}R^{-2}.
\end{array}
\ee
Then, given formula of the variation of the gravity constant (\ref{f18}), the equations 
(\ref{f28}) with conditions (\ref{f31}) are equal
\bea\l{f33}
\dot{N}\dot{R}N^{-5}R^{-1}=8\pi G_{0}\rho_{0}(t)\\
\l{f34}
-c_{0}^2\dot{R}\dot{N}R^{-1}N^{-5}=8\pi G_{0}p_{0int}\\
\l{f35}
c_{0}^{2}\ddot{R}N^{-4}R^{-1}=8\pi G_{0}p_{0ext} \ .
\eea

Let us denote
\be\l{f36}
K=8\pi G_{0}\rho_0
\ee
and find several solutions (\ref{f35}), which have physical meaning for the case of   possible 
generation of matter $\rho_{0}(t)=\rho_{0}/N(t)^d.$ Let's assume that 
\be
N=R \ .
\ee
Then, given generation of matter, the equation (\ref{f33}) transforms into form 
$\dot{N}^2N^{2D}=K,$ where $D=d/2-3.$ If $D\neq{1},$ its solution is
\be
N(t)=\left[1-(D+1)\sqrt{K}(t-t_0)\right]^{1/(D+1)}.
\ee
Denoting $\hat{t}=t_0-t$ we obtain
\be\l{f37}
\hat{N}(\hat{t})\equiv N(t(\hat{t}))=\left[1-(D+1)\sqrt{K}\hat{t}\right]^{1/(D+1)}.
\ee
Now we change over to coordinate system $(t_0,O)$ by denoting 
\be
\hat{\tau}=\int_{0}^{\hat{t}}\hat{N}(\hat{t})d\hat{t}.
\ee
Then (\ref{f37}) is rewritten as
\be\l{f38}
N_{\tau}(\hat{\tau})\equiv 
\hat{N}(\hat{t}(\hat{\tau}))=\left[1+(D+2)\sqrt{K}\hat{\tau}\right]^{1/(D+2)}.
\ee

\subsection{Cosmological parameters}
\l{sec8}

The mean value of density of matter in the Universe, obtained by different techniques 
\cite{bahc} with error $50-75\%$, comprises about 1/5 of critical density, defined with help 
of the Fridman-Robertson-Walker model, $\rho_{crit}=3H^2/(8\pi G_0).$ The field 
equation (\ref{f33}) gives at $t=t_0$: 
\be
8\pi G_0\rho_0=\dot{N}(t_0)\dot{R}(t_0)
\ee
or in view of (\ref{f11}):
\be
8\pi G_0\rho_0=-H\dot{R}(t_0).
\ee
Assuming that 
\be\l{f40}
\dot{N}(t_0)=\dot{R}(t_0)
\ee
we obtain the density value
\be\l{f41}
\rho_0=\frac{1}{8}\frac{H^2}{\pi G_0}.
\ee
which is within the limits of the density of matter in the Universe, derived from 
measurements. It is natural to assume,  that if equality of rate of change of the time scale 
factor and length of internal space (\ref{f40}) takes place at present,  it is fulfilled 
permanently. 

Let us determine dependence between the magnitude of redshift and the distance. The 
distance to the light source is $r=c_0\hat{\tau}.$ From (\ref{f36}) and (\ref{f41}) follows 
that $K=H^2.$ Then, in view of  (\ref{f10}) and (\ref{f37}), (\ref{f38}) we obtain
\be
z=\left[1+(D+2)\frac{Hr}{c_0}\right]^{1/(D+2)}-1.
\ee
If $D>-1,$ then this result gives the picture to be analogous to accelerated expansion in the 
expanding Universe model.

\section{CONCLUSION}  

Value of density of matter in the Universe, determined with help of cosmological model in 
five dimensions with length scale factor which is constant in external space and changing in 
internal space and variable time scale factor, is in agreement with observation data. An 
apparent additional acceleration, indicated from Pioneer 10/11 and Ulysses data, does not 
run counter to this model.  At the same time, conclusion about energy emission and 
generation of matter as a result of decrease of time scale factor needs additional experimental 
confirmation. 

Thus, these results testify about possibility of length time factor to be stationary in the 
significant part of the observed Universe. 

\acknowledgments

 The author would like to thank A.I. Cigan of the Ioffe Physical-Technical Institute for 
valuable discussion.


\begin{references}

\bibitem{milne1} E.A. Milne, {\em Relativity, Gravitation and World-Structure\/}
(Oxford University Press, Oxford, 1935).

\bibitem{milne2} E.A. Milne, Proc. R. Soc. London Ser. A {\bf 158}, 324 (1937).

\bibitem{dirac} P.A.M. Dirac, Nature {\bf 139}, 323 (1937).

\bibitem{bahc} N.A. Bahcall and X. Fan, PNAS {\bf 95}, 5956 (1998) 
(also astro-ph/9804082).

\bibitem{overd} J.M. Overduin and P.S. Wesson, Phus. Rept. {\bf 283}, 303 (1997)
(also gr-qc/9805018).

\bibitem{wesson} P.S. Wesson and H. Liu, Aph. J. {\bf 440}, 1 (1995).

\bibitem{kolb} E.W. Kolb, M.J. Perry and T.P. Walker, Phys. Rev. D{\bf 33}, 869 (1986).

\bibitem{barrow} J.D. Barrow, Phys. Rev. D{\bf 35}, 1805 (1987).  

\bibitem{land} L.D. Landau, E.M. Lifshitz, {\em Teoriya Polja\/} 
(Nouka, Moscow, 1973).

\bibitem{good} S.P. Goodwin, J. Gribbin and M.A. Hedry, Astron. J. {\bf 114}, 2212 
(1997). 

\bibitem{sand} A. Sandage, Astron. J. {\bf 111}, 18 (1996).

\bibitem{vittie} G.C. McVittie, {\em General Relativity and Cosmology\/}
(Chapman and Half Ltd., London, 1956).

\bibitem{bondi} H. Bondi, {\em Cosmology\/}
(Cambridge University Press, Cambridge, 1961).

\bibitem{hoyle} F. Hoyle, Nature {\bf 163}, 4136 (1949). 

\bibitem{jordan} P. Jordan, {\em Schwerkraft und Weltall\/} 
(Fr. Vieweg und Sohn, Braunschweig, 1952).

\bibitem{geoph} I.N. Yanitscy, in: {\em Geophysica i Sovremenny Mir, Megdunarodnaya 
Nouchnaya Conferenciya, Doklady i Referaty\/} (VINITI Publ., Moscow, 1993).

\bibitem{socol} K.I. Sokolovsky, K.E. Yesipchuk, in: {\em Geophysica i Sovremenny Mir, 
Megdunarodnaya Nouchnaya Conferenciya, Doklady i Referaty\/} (VINITI Publ., Moscow, 
1993).

\bibitem{anders} J.D. Anderson et al., Phys. Rev. Lett. {\bf 81}, 2858 (1998) 
(also gr-qc/9808081).

\bibitem{reph} Y. Rephaeli and D. Yankovitch, Astrophus. J. {\bf 481}, L55 (1997).

\bibitem{lasen} A.N. Lasenby and M.E. Jones, in: {\em The Extragalctic Distance Scal\/}
 edited by M. Livio, M. Donahue and N. Panagia (Cambridge University Press, Cambridge, 
1997).

\bibitem{einst} A. Einstein, {\em The Meaning of Relativity\/}
(Princeton University Press, Princeton,1953).

\end{references}
\end{document}